\documentclass[11pt]{article}
\usepackage{moriond,epsfig}

\def\be{\begin{equation}}
\def\ee{\end{equation}}
\def\bea{\begin{eqnarray}}
\def\eea{\end{eqnarray}}

\begin{document}
\vspace*{4cm}
\title{EXCLUSIVE \boldmath{$J/\Psi$} PRODUCTION IN $PP$ AND $P\bar{P}$ COLLISIONS\\ 
AND THE QCD ODDERON}

\author{ ADAM BZDAK \footnote{Fellow of the Polish Science Foundation (FNP) scholarship
for the year 2006}}

\address{M. Smoluchowski Institute of Physics, Jagiellonian University,\\
Reymonta 4, 30-059 Krak\'{o}w, Poland}

\maketitle\abstracts{We briefly present the QCD based calculation of exclusive $J/\psi$ production
via pomeron-odderon fusion in proton-antiproton collision. For the Tevatron
energy the differential cross section $d\sigma/dy(y=0)$ is estimated to be
around $1$ nb.}

\section{Introduction}

The color neutral gluon systems, exchanged at high energy scattering
processes, can be classified with respect to their $C$ parity. The most
important one is $C$-even system with quantum numbers of vacuum $i.e.$ the
pomeron. In perturbative QCD the lowest order prototype of the pomeron is the
color neutral system of two gluons. The odderon is the $C$-odd partner of the
pomeron. The hard odderon skeleton consists of three gluons in a color neutral
state. It is quite obvious that the effects of the odderon exchanges are less
important than those due to the pomeron (one would naively expect a
suppression by a power of the coupling constant $\alpha_{s}$ for the
additional gluon). It is not clear, however, why the contribution of the
odderon is so small that it has not been definitely observed by any experiment.

The concept of the odderon was introduced a long time ago \cite{Lukaszuk}. To
this day the best, but still weak, experimental evidence for the odderon was
found as a difference between the differential cross sections for elastic $pp$
and $p\bar{p}$ scattering at $\sqrt{s}=53$ GeV at CERN ISR \cite{Breakstone}.
A natural difficulty in detecting odderon effects is the fact that, in
general, the odderon and the pomeron contribute to the scattering amplitude at
the same time. For a detailed review of the phenomenological and theoretical
status of the odderon we refer the reader to \cite{Ewerz-review}.

Recently more attention has been given to exclusive production processes in
which the odderon is the only possible exchange. In the present paper we
concentrate on exclusive $J/\psi$ meson production in $p\bar{p}$ collisions:
$p\bar{p}\rightarrow p+J/\psi+\bar{p}$ where ''$+$'' means rapidity gap. This
process occurs via pure odderon exchange without any pomeron mixing. The
odderon is here in competition only with the photon which is under good
theoretical control.

\section{Exclusive $J/\psi$ production in perturbative QCD}

Diffractive production of $J/\psi$ meson in proton-(anti)proton collisions via
pomeron-odderon fusion was investigated in ref. \cite{Schafer} in the
framework of Regge theory. The potential contribution of the $\omega$ reggeon
to this process is expected to be strongly suppressed due to the Zweig rule.
It was estimated that for $p\bar{p}$ collisions at $\sqrt{s}=2$ TeV a total
cross section is of the order $\sigma_{tot}=75$ nb.\footnote{This result does
not take the pomeron-photon fusion contribution into account} This result is
quite encouraging, however, it should be treated rather as a order of
magnitude estimate.

There is another very attractive theoretical feature of this process, that is
a presence of relatively large scale $m_{J/\psi}^{2}$ that, with a bit of
optimism, justifies the use of perturbative QCD. \ 

Now let us step by step explain the main points of our calculation.

\begin{figure}[h]
\begin{center}
\includegraphics[scale=1]{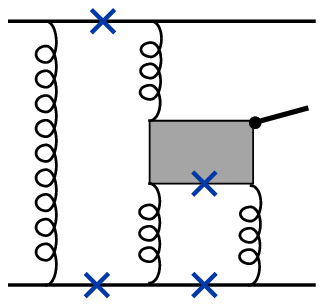}  \hspace{0.5cm}
\includegraphics[scale=1]{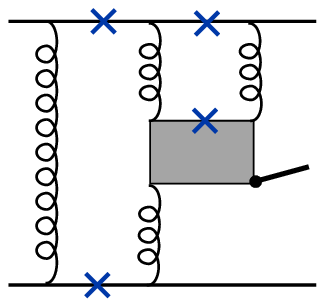}  \hspace{0.5cm}
\includegraphics[scale=1]{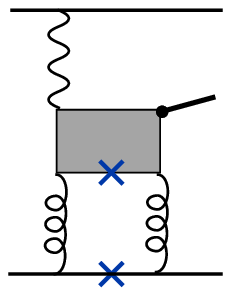}  \hspace{0.5cm}
\includegraphics[scale=1,trim=0 -2 0 0]{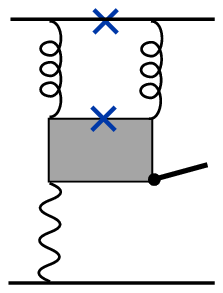}
\end{center}
\caption{Sixteen (see Fig. \ref{M_a}) diagrams contributing to the amplitude
of the process of exclusive $J/\psi$ meson production via pomeron-odderon and
pomeron-photon fusion in quark-(anti)quark collision. }%
\label{M_1}%
\end{figure}

\begin{figure}[h]
\begin{center}
\includegraphics[scale=1]{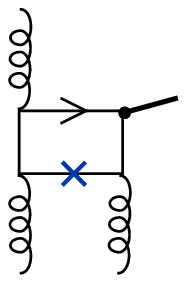}  \hspace{0.4cm}
\includegraphics[scale=1]{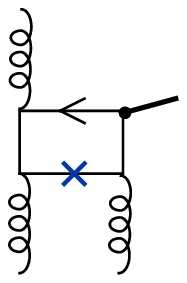}  \hspace{0.4cm}
\includegraphics[scale=1]{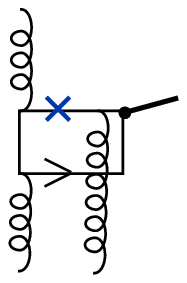}  \hspace{0.4cm}
\includegraphics[scale=1]{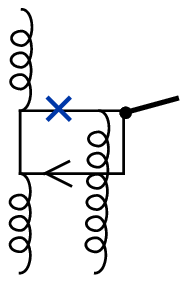}
\end{center}
\caption{The shaded area in Fig. \ref{M_1} represents a sum of four diagrams
contributing to the subprocess $3g\rightarrow J/\psi$.}%
\label{M_a}%
\end{figure}

The Born amplitude \footnote{The details of our calculations will be presented
in a forthcoming paper} (multiplied by $2!$) of exclusive $J/\psi$ production
in quark-(anti)quark collision is described by the sum of sixteen diagrams
shown in Fig. \ref{M_1} and Fig. \ref{M_a} (the quark line with a cross is
put\textit{ on shell}).

The shaded area in Fig. \ref{M_1} denotes a sum of four diagrams shown in Fig.
\ref{M_a} (and analogous for the $\gamma gg\rightarrow J/\psi$ subprocess).

Here some comments are to be in order.

1.The coupling of the pomeron and the odderon to the proton. The model we
adopt was formulated by Fukugita and Kwieci\'{n}ski \cite{Fukugita}. Shortly
speaking, we consider the proton as a system of three valence quarks with
totally antisymmetric wave function in the color space. The strong coupling
constant is taken to be $1$, however, the real value may be as small as $0.3$
\cite{Ewerz-coupling}. It is the main source of uncertainties in our calculation.

2.The $J/\psi$ vertex. As the binding energy of $s$ wave $c\bar{c}$ quarks for
$J/\psi$ system is small (much less than the charm quark mass), we can follow
\cite{Berger,Ryskin} and use nonrelativistic colinear approximation. The
coupling constant may be expressed in terms of the electronic width
$\Gamma_{e^{-}e^{+}}^{J/\psi}$ of $J/\psi\rightarrow e^{-}e^{+}$ decay. It was
suggested in \cite{relativistic} to implement relativistic corrections by
redefining the mass of the $c$ quark to be half of the $J/\psi$ mass.

3.Gap survival factor. We also must take the gap survival effect into account
$i.e.$ the probability $S_{gap}^{2}$ of the gaps not to be populated by
secondaries produced in the soft rescattering. It is not a universal number
but it depends on the initial energy and the particular final state. In our
calculations we take for the Tevatron energy for pomeron-odderon fusion
contribution $S_{gap}^{2}=0.05$ \cite{KMR-chi}. Here we assume that the soft
rescattering in exclusive $J/\psi$ production via pomeron-odderon fusion is the
same as in the reaction of exclusive double pomeron exchange $\chi_{c}$
production (actually it is the best we can do). For the pomeron-photon fusion
contribution we take $S_{gap}^{2}=1$ \cite{KMR-photon}.

4.BFKL evolution \cite{BFKL}. For the Tevatron energy we expect enhancement by
a factor about $3$.

5.BKP evolution \cite{BKP}. At leading order the intercept of the odderon
trajectory is predicted to be very close to $1$ \cite{Wosiek}. From this point
of view bare three gluons in color neutral state appear as a natural model for
the odderon.

\section{Results and discussion}

Now we are ready to show our predictions for exclusive $J/\psi$ meson
production in proton-(anti)proton collisions.

Let us start from some general remarks. The cross section in proton-proton
collisions is smaller than in proton-antiproton collisions. The differential
cross section $d\sigma/dy$ weakly depends on the rapidity $y$ of the produced
$J/\psi$. It is presumably a reflection of lack of BFKL and BKP evolution.
Another interesting observation is that the odderon and photon contributions
do not interfere (complex phase). It allows us to discuss these contributions separately.

In Table \ref{results} predictions for the Tevatron energy are
shown.\footnote{Note that the branching fraction for the decay $J/\psi
\rightarrow\mu^{+}\mu^{-}$ is not included} As can be seen the
pomeron-odderon contribution to the differential cross section for exclusive
$J/\psi$ meson production is of comparable size as the contribution coming
from pomeron-photon fusion. It should be noted, however, that the odderon
coupling to proton is the most serious uncertainty of our calculations and the
result presented in Table \ref{results} for pomeron-odderon contribution
should be regarded only as an order of magnitude estimate.

\begin{table}[h]
\begin{center}%
\begin{tabular}
[c]{|c|c|c|}\hline\hline
$\sqrt{s}=2$ TeV & odderon & photon\\\hline
$d\sigma/dy(y=0)$ & $0.5-3$ nb & $2.5$ nb\\\hline
\end{tabular}
\end{center}
\caption{The results for exclusive $J/\psi$ production at the Tevatron energy
in proton-antiproton collisions based on Fukugita-Kwieci\'{n}ski model. The
main uncertainty comes from unknown odderon-proton coupling.}%
\label{results}%
\end{table}

The situation looks much better if we impose the following cuts: $\left|
t_{1}^{2}\right|  ,$ $\left|  t_{2}^{2}\right|  >0.25$ GeV$^{2}$ (or
appropriate cuts for one hadron and transverse momentum of $J/\psi$). Then,
the pomeron-odderon fusion contribution decreases about an order of magnitude,
being still visible, and the pomeron-photon fusion contribution decreases
about two to three orders of magnitude, being completely negligible.

At the end one thing should be emphasized. The cross section of $\gamma
p\rightarrow J/\psi+p$ subprocess may be calculated perturbatively (as we
did), see for example \cite{Ryskin}, or taken from extrapolations of the HERA
data \cite{HERA}. It allows us to calculate the pomeron-photon fusion
contribution in a model independent way. Performing appropriate calculations
we obtain the value to be about $3$ nb, what is in good agreement with our
perturbative calculation. For details of this approach see \cite{KMR-photon}
where this problem is thoroughly studied.\footnote{In this reference the
authors also estimate the pomeron-odderon contribution. However, they use
different diagrams -- the whole odderon is coupled to $c$ quark} We believe
that it allows to estimate this contribution with an accuracy up to $15\%$.

In summary, our recipe for the odderon looks like follows: take the data for
exclusive $J/\psi$ production, throw away the pomeron-photon contribution
being under quite good control, and if there remains something, and we suggest
that this will be indeed the case, it is the odderon.

\section*{Acknowledgments}
This work was done in collaboration with: J. R. Cudell, L. Motyka and L.
Szymanowski. The author would like to thank the organizers of the 41th
Rencontres de Moriond for support.

\end{document}